\documentclass[11pt, oneside,reqno]{amsart}

\usepackage[top=2.25cm, bottom=2.25cm, left=2.25cm, right=2.25cm]{geometry}

\usepackage[english]{babel}

\usepackage[T1]{fontenc}

\usepackage{textcomp}

\usepackage[utf8]{inputenc}

\usepackage{amsfonts, amsmath, amssymb, mathtools, amsthm}
\usepackage[makeroom]{cancel}
\usepackage[none]{hyphenat}
\usepackage{bigints}
\usepackage{xcolor}

\usepackage{siunitx}
\usepackage{fancyhdr}
\usepackage{graphicx}
\usepackage{verbatim}
\usepackage{xcolor}
\usepackage{pgf, tikz, tkz-base}
\usepackage{mathrsfs}
\usepackage{tkz-euclide}
\usepackage{xfp}

\usetikzlibrary{shapes, calc, shapes, arrows, babel}
\usepackage[margin=16pt,font=small,labelfont=bf]{caption} 

\usepackage[maxbibnames=99, sorting=nyt]{biblatex}
\usepackage{parskip}

\usepackage[colorlinks=true]{hyperref}

\usepackage{csquotes}
\usepackage{biblatex}

\usepackage{fancyhdr}
\begin{document}

\title
{Spin orbit resonance cascade via core shell model.\\
Application to Mercury and Ganymede}

\author{
Gabriella Pinzari$^{1}$, 
Benedetto Scoppola$^{2}$, 
Matteo Veglianti$^{3}$}


\maketitle

\begin{center}
{
\footnotesize
\vspace{0.3cm}$^{1}$  Dipartimento di Matematica “Tullio Levi–Civita”,\\ Università degli Studi di Padova,
\\Via Trieste, 63, 35131 Padova, Italy\\
\texttt{Gabriella.Pinzari@math.unipd.it}\\

\vspace{0.3cm}$^{2}$  Dipartimento di Matematica,\\
Universit\`a di Roma
``Tor Vergata''\\
Via della Ricerca Scientifica - 00133 Roma, Italy\\
\texttt{scoppola@mat.uniroma2.it}\\

\vspace{0.3cm} $^{3}$ Dipartimento di  Matematica,\\ Universit\`a di Roma
``Tor Vergata''\\
Via della Ricerca Scientifica - 00133 Roma, Italy\\
\texttt{veglianti@mat.uniroma2.it}\\ 
}

\end{center}

\begin{abstract}
We discuss a model describing the spin orbit resonance cascade.
We assume that the primary has a two-layer (core-shell) structure: it is composed by a thin solid crust and an inner and heavier solid core that are interacting due to the presence of a fluid interface.
We assume two sources of dissipation: a viscous one, depending on the relative angular velocity between core and crust and a tidal one, smaller than the first, due to the viscoelastic structure of the core.
We show how these two sources of dissipation are needful for the capture in spin-orbit resonance. The crust and the core fall in resonance with different time scales if the viscous coupling between them is big enough.
Finally, the tidal dissipation of the viscoelastic core, decreasing the eccentricity, brings the system out of the resonance in a third very long time scale. This mechanism of entry and exit from resonance ends in the $1:1$ stable state.
\end{abstract}

\section{Introduction}
It is well known that celestial mechanics is a very effective way to describe the astronomical motions because the systems are in this context almost conservative. However considering the fact that the bodies are extended, instead of point masses, and that their inner motions, mainly due to tides, dissipate energy, some small non conservative effects has to be taken into account. More recently the argument has been deeply studied, and it became very important also because we have today the possibility to perform astronomical measurements with an impressive precision. Actually, the main motivation of this work is to introduce an interpretative framework for the measurement on Ganymede of the space mission JUICE. Recent space missions (JUICE, Juno, BepiColombo), indeed, are receiving considerable support frameworks in literature, consisting in both analytically and numerically models, see, for instance \cite{lari2022orbit, lari2018semi}.\\
The literature about tidal dissipation and triaxial effects is immense, see for instance \cite{MurrayDermott, efroimsky2013tidal, efroimsky2015tidal, ferraz2015dissipative} and references therein. The subject is quite subtle, most of all because the friction appearing in this dissipation is a very complicated phenomenon, and we have so far only phenomenological models of it. Just to quote a relatively recent development with this respect, in a geological framework the motion between plaques is supposed to exhibit the so-called stick/slip phenomenon, see for instance \cite{corbi}, which surely goes beyond the possibility of a detailed control of the parameters of the system.\\
Some attempt, in a statistical mechanics context, has been preliminary performed in terms of the so-called shaken dynamics, see for instance \cite{scoppola2022shaken, apollonio2022shaken, apollonio2022metastability, d2021parallel, apollonio2019criticality}, but the subject is really in a very primordial state.
On the other side a quantitative control of the friction involved in the tidal phenomena could be very useful in order to study the numerous resonances that can be observed in solar system. While it is definitely quite clear that such resonances may appear in celestial mechanics, see again for instance \cite{MurrayDermott, correia2018effects, antognini2014spin} and, in terms of a simplified model, \cite{scoppola2022tides}, it is much less clear, from a quantitative point of view, how it is possible that the systems are captured by certain resonances. An example particularly clear with this respect is the 3:2 spin-orbit resonance exhibited by Mercury. The latter has been investigated in pioneering works, see \cite{goldreich1966spin, goldreich1967spin}, and it has been immediately clear that the probability of capture in the observed resonance for Mercury is extremely small if one wants to describe the tidal friction in terms of simple phenomenological relations. After this first computation, various models have been proposed in order to overcome this difficulty, giving a more plausible justification of the observed resonance. Here we want to recall the paper \cite{noyelles2014spin}, in which the capture in resonance has been described in term of a very detailed and frequency dependent model of tidal friction. On the other hand, numerical approaches are also widely studied, see, for instance \cite{bartuccelli2015high, correia2009mercury, correia2010long}.\\
In this paper we propose a slightly different description of the system, based on some basic hypothesis about the inner structure of the primary, that seems to be quite accepted in literature, and on simple laws of friction, similar to the ones appearing in \cite{goldreich1966spin, goldreich1967spin}. The basic idea is the following: assume that the primary is composed by a thin solid crust and an inner and heavier solid core, both triaxial, that are interacting due to the presence of a fluid interface. This interaction, when the rotations of the crust and the core are different, is a first source of dissipation. A second, much smaller, source of friction is due to the viscoelastic structure of the core, in which the tidal deformation are not completely elastic, and therefore imply a certain dissipation.
We will face in the rest of the paper two different systems, namely Sun-Mercury and Jupiter-Ganymede. The structure described above is shared by these two systems, with an important difference: indeed, according to Juno's data, the liquid interface in the case of Ganymede is supposed to be composed of liquid water, see for instance \cite{gomez2022gravity}; Mercury, on the contrary has probably a molten mantle, exhibiting a viscosity about $7$ order of magnitude higher than the water's, see for instance \cite{smith2012gravity}.
In literature it is possible to find some studies, related to different phenomena, starting from these assumptions on the inner structures of the celestial bodies, see for instance \cite{folonier2017tidal, baland2019coupling, ragazzo2022librations}.\\
In order to describe the two sources of dissipation in the system we assume a simple viscous friction, linearly dependent on the difference of velocity, for the interaction between crust and core. As it will be clear below, using a rough computation based on laminar solution of the Navier-Stokes equation, this assumption seems to be reasonable, and it is possible to give an estimate of the order of magnitude of the torque produced by this kind of friction. The description of the capture in resonance, anyway, is not dependent on this linearity assumption: we just need a continuous friction vanishing at zero velocity.
The details about the viscoelastic friction on the core are even less important in order to achieve the results of this paper. The only important assumption with this respect is the fact that the viscoelastic tidal torque on the core has to be smaller than the viscous torque applied to the crust-core system. Rough estimates, see again next sections, show that this assumption is fulfilled in the two specific systems we are studying.
Since the details of the viscoelastic torque on the core are inessential we will describe it again in terms of a friction linear in velocity, in order to be able to write the equation of motions by means of a Rayleigh function.\\
As it will be clear below, the mechanism leading to the capture in spin-orbit resonance of the celestial bodies seems to be the following. The (much lighter) crust usually rotates jointly with the core. On the latter it is always present a viscoelastic tidal torque, that slowly decreases the angular velocity of the whole body. When the angular velocity of such whole body becomes sufficiently close to a resonance the crust is certainly captured in resonance on a very small time scale. Then, if the contribution of the viscoelastic friction is sufficiently small with respect to the viscous coupling between crust and core, the core is driven, again with certainty, to the same resonance on a different and slightly longer time scale. The body, then, exits from resonance when the action of the viscoelastic torque reduces the eccentricity of the orbit under a certain limit value, and this happens on a third longer time scale. The whole process causes a cascade of spin-orbit resonances, bringing eventually the system in the stable 1:1 state.\\
In this paper for the seek of simplicity we consider only the leading terms of expansions in the small parameters of the system. Future works will be devoted to the control of the convergence of such expansions following the approach presented in \cite{chen2021exponential, celletti2000hamiltonian, calleja2022kam}. 
The work is organized as follows. In section~2, to recall the mechanism leading to the spin-orbit resonance and to fix some notations, we present the equations of motions of the two layers without viscous coupling. Then we evaluate the stability of any generic spin-orbit resonance. In section~3 we introduce the two dissipative terms, we discuss the capture mechanisms and we find out a condition on the eccentricity that ensure the capture into spin-orbit resonance for the crust and the core. In section~4 we show how the system exits from a resonance to go to the next one. Finally, section~5 is devoted to some preliminary numerical computation and to the discussion of future developments of the work.

\section{Stability of spin-orbit resonance}

\begin{figure}
    \label{fig.system}
    \centering
    \includegraphics[width=0.6\textwidth]
        {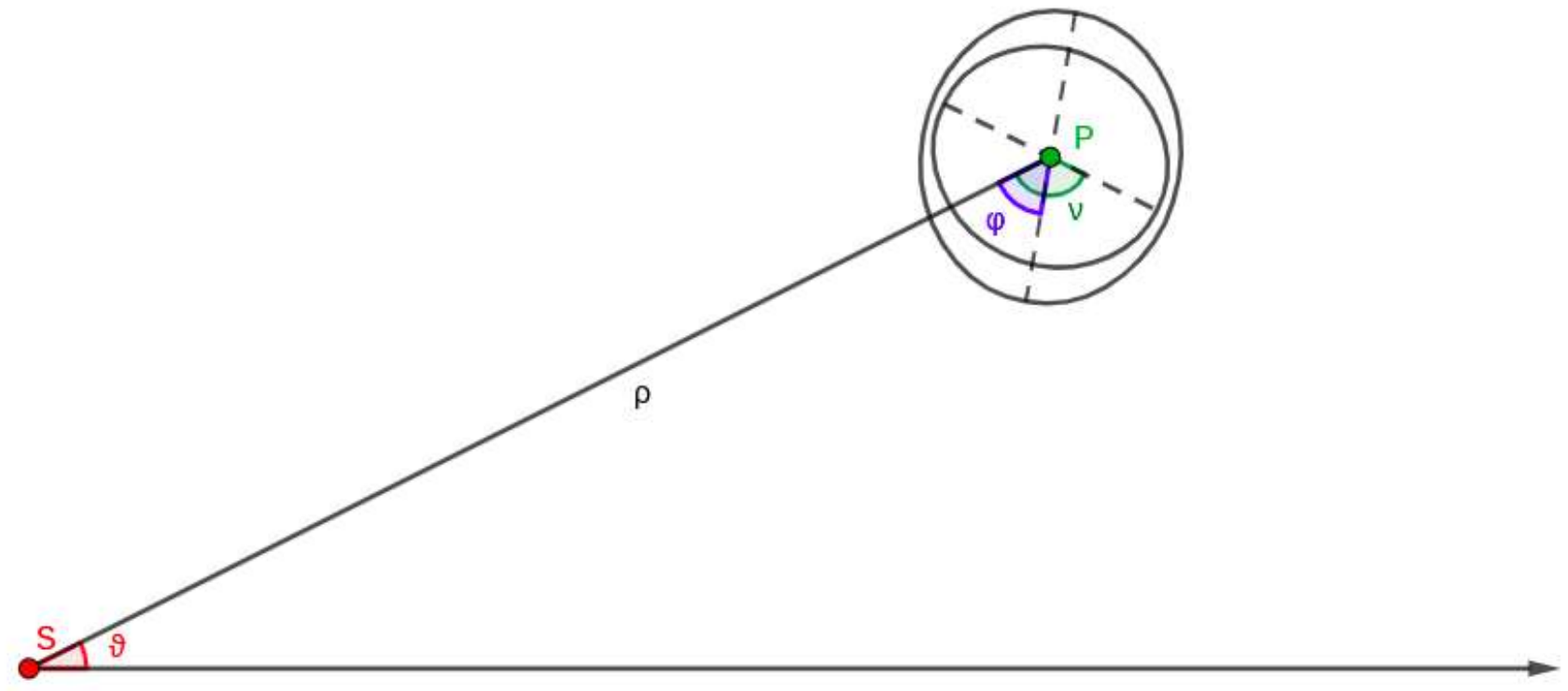}
    \caption{Geometry of the system. The triaxialities of core and crust have been exaggeratedly increased to make them graphically appreciable.}
\end{figure} 

Celestial bodies such as Mercury or some of the Jovian satellites consists of several layers of materials with different density and interacting differently with other celestial body (e.g. the Sun, Jupiter or the other satellites).
The rotational velocity of such layers is, in general, not the same. It is reasonable to think that the friction between them depends on their mutual angular velocity.\\
In this framework we introduce a simple model in order to compute the effects of inner dissipation in the two body problem. We want to investigate the effects that this dissipation has on the orbital evolution and in particular on the mechanism involved in spin-orbit resonance capture. We will assume that the two bodies have very different masses, say $M\gg m$, and we call {\it secondary} the body with mass $M$ and {\it primary} the body with mass $m$. To keep the model as simple as possible we assume the axis of rotation of the primary to be perpendicular to the orbital plane.\\
In our model, we assume a two-layer structure of the primary: an inner solid not perfectly elastic heavy core and an external solid thin and light crust, both described in terms of triaxial ellipsoid, interacting with each other. So the primary, centered in $P$, is described in terms of a not perfectly spherical core of mass $m-\mu$ and a not perfectly spherical crust of mass $\mu$. Let $A,B,C$ and $A',B',C'$ be respectively the moments of inertia of the core and of the crust with respect to the reference axis. In particular, $C$ and $C'$ are the the moments of inertia with respect to the spin direction. The equatorial ellipticity are therefore $\varepsilon = \frac{3}{2}\frac{B-A}{C}$ and $\varepsilon' = \frac{3}{2}\frac{B'-A'}{C'}$ respectively. It is reasonable to assume that the triaxiality shape of core and crust are similar: $\varepsilon \simeq \varepsilon'$; but, since the crust is lighter and thinner than the core, it is also reasonable to assume that each moment of inertia of the crust is smaller than the corresponding moment of inertia of the core, in particular $C'<<C$.\\
The results we obtain remain unchanged if one consider a simpler model in which the primary, centered in $P$, is described in terms of a not perfectly spherical core of mass $m-\mu$ and a mechanical dumbbell centered in $P$, i.e., a system of two points, each having mass $\mu/2$, constrained to be at fixed mutual distance $2l$, having $P$ as center of mass. The idea is to substitute the triaxiality oh the crust with a dumbbell. In this paper, however, we follow the first more usual approach.\\
Let the secondary be a massive point $S$ fixed in the origin of reference frame and let $P$ moves on a Keplerian orbit around $S$. Let $\rho$ be the distance $PS$, $\vartheta$ be the true anomaly and let $\varphi$ and $\nu$ be the angles between the $PS$ direction and direction of the major axis of the crust or the direction of the major axis of the core respectively. So, $\dot \vartheta$, $\dot \varphi$ and $\dot \nu$ are the orbital angular velocity and the angular velocities with respect to the secondary of the crust and the core respectively.\\
In order to be able to insert friction in the system in terms of Rayleigh's dissipation function, we use a Lagrangian Formalism.

The total kinetic energy is the sum of the kinetic energies of the two shells: the crust and the core.
 \begin{equation}
        \mathcal{T} = \frac{1}{2} m (\dot{\rho}^2 + {\rho}^2 \dot{\vartheta}^2) +\frac{1}{2} C'(\dot{\varphi} + \dot{\vartheta})^2 + \frac{1}{2} C (\dot{\nu} + \dot{\vartheta})^2.
\end{equation}
The potential energy is the sum of two pieces of gravitational attraction: the attraction between the core of the primary and the secondary and the attraction between the crust of the primary and the secondary:
\begin{equation}
    \label{eq.potential}
    \mathcal{V} = -GM\int_{V}\frac{dm}{r} -GM\int_{V'}\frac{dm}{r} = -\frac{GMm}{\rho} - \frac{G M (B-A)}{4\rho^3} \left[1 + 3\cos(2\nu) \right] -\frac{G M (B'-A')}{4\rho^3} \left[1 + 3\cos(2\varphi) \right].
\end{equation} 
where $G$ is the universal gravitational constant, $V$ the volume of the core and $V'$ the volume of the crust.\\ 
So, the energy of the system is:
\begin{equation}
    \label{eq.energy}
    E = E_k + E_{\varphi} + E_{\nu},
\end{equation}
with:
\begin{equation}
    \label{eq.energy_k}
    E_k = \frac{1}{2}m(\dot{\rho}^2 + {\rho}^2 \dot{\vartheta}^2) -\frac{GMm}{\rho},
\end{equation}

\begin{equation}
    \label{eq.energy_fi}
    E_{\varphi} = \frac{1}{2} C' (\dot{\varphi} + \dot{\vartheta})^2 - \frac{GM (B'-A')}{4a^3} \left( \frac{a}{\rho} \right)^3 \left[1 + 3\cos(2\varphi) \right].
\end{equation}
and
\begin{equation}
    \label{eq.energy_ni}
    E_{\nu} = \frac{1}{2} C (\dot{\nu} + \dot{\vartheta})^2 - \frac{GM (B-A)}{4a^3} \left( \frac{a}{\rho} \right)^3 \left[1 + 3\cos(2\nu) \right].
\end{equation}

We note that $E_k$ is the well known energy of the Kepler problem, that is constant on the Keplerian orbit, describing the motion of the center of mass of the primary.\\
In what follows we assume that the center of mass of the primary moves unperturbed on its Keplerian trajectory.\\
On the other hand $E_{\varphi}$ and $E_{\nu}$ describes the dynamics associated to the crust and the core respectively.\\
It is useful to consider the entire series of $(a/\rho)^3$ and $\vartheta$ in terms of eccentricity and mean anomaly, see \cite{MurrayDermott}, namely:
\begin{equation}
    \label{eq.rho}
    \left( \frac{a}{\rho} \right)^3 = \sum_{n=0}^{\infty} a_n(e) e^n \cos(n\omega t),
\end{equation}
with
\begin{equation}
    \label{eq.a_coeff}
    a_n(e) = \sum_{l=0}^{\infty} b_{nl} e^{2l}, 
\end{equation}
and $b_{00} = 1$.

\begin{equation}
    \label{eq.theta}
    \vartheta = \omega t + \sum_{n=1}^{\infty} c_n(e) e^n \sin(n\omega t),
\end{equation}
with
\begin{equation}
    \label{eq.c_coeff}
    c_n(e) = \sum_{l=0}^{\infty} d_{nl} e^{2l}.
\end{equation}
Hence:
\begin{equation}
    \label{eq.theta_dot}
    \dot{\vartheta} = \omega + \sum_{n=1}^{\infty} n \omega c_n(e) e^n \cos(n\omega t).
\end{equation}

So $E_{\varphi}$ and $E_{\nu}$ can be written as (cfr. \eqref{eq.rho}, \eqref{eq.theta_dot}):
\begin{equation}
    \label{eq.E_prime}
        E_{\varphi} = \frac{1}{2} C' \left(\dot{\varphi} + \omega + \sum_{n=1}^{\infty} n \omega c_n(e) e^n \cos(n\omega t) \right)^2 - (B'-A') \frac{\omega^2}{4} \left[1 + 3\cos(2\varphi) \right] \sum_{n=0}^{\infty} a_n(e) e^n \cos(n\omega t)
\end{equation}
and
\begin{equation}
    \label{eq.E_prime_nu}
        E_{\nu} = \frac{1}{2} C \left(\dot{\nu} + \omega + \sum_{n=1}^{\infty} n \omega c_n(e) e^n \cos(n\omega t) \right)^2 - (B-A) \frac{\omega^2}{4} \left[1 + 3\cos(2\nu) \right] \sum_{n=0}^{\infty} a_n(e) e^n \cos(n\omega t),
\end{equation}
where, we have used the Kepler's third law: $\frac{GM}{a^3} = \omega^2$.

We can now consider a generic $2(k/2 +1) : 2$ spin-orbit resonance, setting
\begin{equation}
    \label{eq.phi_resonant}
        \varphi = \frac{k \omega t}{2} + \gamma 
        \footnote{
            Indeed, according to our definition, the angle between the semi-major axis of the crust and the $x$-axis is $\vartheta + \varphi \simeq \omega t + \frac{k \omega t}{2} + \gamma = \left( \frac{k}{2} + 1 \right) \omega t + \gamma$; so the crust's angular speed with respect to the $x$-axis is $\left( \frac{k}{2} + 1 \right)\omega  + \dot{\gamma}$, that is the $2(k/2 +1) : 2$ spin-orbit resonance condition, with $\gamma$ the resonant angle. Same for the core.}
        \text{ \ \ \ and \ \ \ } 
        \nu = \frac{k \omega t}{2} + \eta.
\end{equation}
Hence \eqref{eq.E_prime} and \eqref{eq.E_prime_nu} become:
\begin{equation}
    \label{eq.E_prime_resonant}
     E_{\gamma} = \frac{1}{2} C' \left[ \left(\frac{k}{2}+1 \right)\omega + \dot{\gamma} + \sum_{n=1}^{\infty} n \omega c_n(e) e^n \cos(n\omega t) \right]^2 - (B'-A') \frac{\omega^2}{4} \left[1 + 3\cos(k \omega t + 2\gamma) \right] \sum_{n=0}^{\infty} a_n(e) e^n \cos(n\omega t)
\end{equation}
and
\begin{equation}
    \label{eq.E_prime_eta_resonant}
     E_{\eta} = \frac{1}{2} C \left[ \left(\frac{k}{2}+1 \right)\omega + \dot{\eta} + \sum_{n=1}^{\infty} n \omega c_n(e) e^n \cos(n\omega t) \right]^2 - (B-A) \frac{\omega^2}{4} \left[1 + 3\cos(k \omega t + 2\eta) \right] \sum_{n=0}^{\infty} a_n(e) e^n \cos(n\omega t).
\end{equation}

Each of these expressions depend on two angles $\vartheta = \omega t$ and $\gamma$ or $\eta$ respectively. According to our assumptions, $\vartheta$ is faster than $\gamma$ and $\eta$ (that vary very slowly), for this reason we can consider the mean value of the energies $E_{\gamma}$ and $E_{\eta}$ over a period of $\vartheta$.
\begin{equation}
    \label{eq.energy_prime_mean}
    \langle E_{\gamma}\rangle = \frac{1}{2} C' \left[ \dot{\gamma}^2 + \left(\frac{k}{2}+1 \right)^2 \omega^2 + 2\left(\frac{k}{2}+1 \right)\omega \dot{\gamma} + \sum_{n=1}^{\infty} \frac{n^2 \omega^2 c^2_n(e) e^n}{2} \right] -(B'-A') \frac{\omega^2}{4} \left[1 + \frac{3}{2}\cos(2\gamma)a_k(e)e^k \right]
\end{equation}
and
\begin{equation}
    \label{eq.energy_prime_mean_eta}
    \langle E_{\eta}\rangle = \frac{1}{2} C \left[ \dot{\eta}^2 + \left(\frac{k}{2}+1 \right)^2 \omega^2 + 2\left(\frac{k}{2}+1 \right)\omega \dot{\eta} + \sum_{n=1}^{\infty} \frac{n^2 \omega^2 c^2_n(e) e^n}{2} \right] - (B-A) \frac{\omega^2}{4} \left[1 + \frac{3}{2}\cos(2\eta)a_k(e)e^k \right].
\end{equation}
The corresponding Lagrangian is:
\begin{equation*}
    \mathcal{L} = \frac{1}{2} C' \left[ \dot{\gamma}^2 + \left(\frac{k}{2}+1 \right)^2 \omega^2 + 2\left(\frac{k}{2}+1 \right)\omega \dot{\gamma} + \sum_{n=1}^{\infty} \frac{n^2 \omega^2 c^2_n(e) e^n}{2} \right] +(B'-A') \frac{\omega^2}{4} \left[1 + \frac{3}{2}\cos(2\gamma)a_k(e)e^k \right] + 
\end{equation*}
\begin{equation}
    \label{Lagrangian}
    + \frac{1}{2} C \left[ \dot{\eta}^2 + \left(\frac{k}{2}+1 \right)^2 \omega^2 + 2\left(\frac{k}{2}+1 \right)\omega \dot{\eta} + \sum_{n=1}^{\infty} \frac{n^2 \omega^2 c^2_n(e) e^n}{2} \right] + (B-A) \frac{\omega^2}{4} \left[1 + \frac{3}{2}\cos(2\eta)a_k(e)e^k \right].
\end{equation}
The corresponding equations of motion are:
\begin{equation}
    \label{eq_2equation}
    \begin{cases}
         C' \Ddot{\gamma} = - \frac{3}{4} (B'-A') \omega^2 a_k(e)e^k \sin{2 \gamma}\\
          C \Ddot{\eta} = - \frac{3}{4} (B-A) \omega^2 a_k(e)e^k \sin{2\eta}
    \end{cases}
\end{equation}
that are the equations of two independent pendulums.\\
So, for suitable initial conditions, $\gamma = 0$, $\eta = 0$  is a stable equilibrium point for the system.\\
This implies that every $2(k/2 +1) : 2$ spin-orbit resonance is stable.

\section{Capture into spin-orbit resonance} 

In this section we want to investigate the resonance capture mechanism. In order to do that we introduce in both the equations in \eqref{eq_2equation} a coupled viscous friction (i.e. a friction proportional to the difference of velocity $\dot\varphi - \dot\nu = \dot\gamma -\dot\eta $). Moreover, in equation of $\eta$ we introduce a second term of friction, proportional to $\dot\nu = \dot\eta + \frac{k\omega}{2}$ that gives the dissipation due to the non perfect elasticity of the core.\\
A standard approach to treat a viscous friction in Lagrangian formalism is to use the Rayleigh's dissipation function $R$, defined as the function such that $\frac{\partial R}{\partial \dot{q}_i} = f_i$, where $f_i$ is the frictional force acting on the $i$-th variable.\\
In our case, the Rayleigh's dissipation function assumes the form:
\begin{equation}
    \label{Rayleigh}
    R=-\frac{1}{2}\lambda \left( \dot{\varphi} - \dot{\nu} \right)^2 - \frac{1}{2}\lambda' \dot{\nu}^2 = -\frac{1}{2}\lambda \left( \dot{\gamma} - \dot{\eta} \right)^2 - \frac{1}{2}\lambda' (\dot{\eta} + \frac{k\omega}{2})^2,
\end{equation}
with $\lambda > 0$ a viscous friction coefficient and $\lambda' << \lambda$ a viscoelastic friction coefficient.\\
The Euler-Lagrange equations become:
\begin{equation}
    \label{E_L_eq}
    \begin{cases}
        \frac{d}{dt} \left( \frac{\partial \mathcal{L}}{\partial \dot\gamma} \right) = \frac{\partial \mathcal{L}}{\partial \gamma} + \frac{\partial R}{\partial \dot\gamma}\\
        \frac{d}{dt} \left( \frac{\partial \mathcal{L}}{\partial \dot\eta} \right) = \frac{\partial \mathcal{L}}{\partial \eta} + \frac{\partial R}{\partial \dot\eta}
    \end{cases}
\end{equation}
Then the equations of motion are:
\begin{equation}
    \label{eq_2equation_coupled}
    \begin{cases}
         C' \Ddot{\gamma} = - \frac{3}{4} (B'-A') \omega^2 a_k(e)e^k \sin{2 \gamma} - \lambda (\dot\gamma - \dot\eta)\\
          C \Ddot{\eta} = - \frac{3}{4} (B-A) \omega^2 a_k(e)e^k \sin{2\eta} + \lambda (\dot\gamma - \dot\eta) -\lambda' (\dot\eta + \frac{k\omega}{2})
    \end{cases}
\end{equation}
These equations can be rewritten in terms of first order ones:
\begin{equation}
    \label{eq.velocity_field}
        \begin{cases}
            C' \dot{v_{\gamma}} = - \frac{3}{4} (B'-A') \omega^2 a_k(e)e^k \sin{2\gamma} - \lambda (v_{\gamma} -v_{\eta})\\
            \dot{\gamma} = v_{\gamma}\\
            C \dot{v_{\eta}} = - \frac{3}{4} (B-A) \omega^2 a_k(e)e^k \sin{2\eta} + \lambda (v_{\gamma} - v_{\eta}) -\lambda' (v_{\eta} + \frac{k\omega}{2})\\
            \dot{\eta} = v_{\eta}
        \end{cases}
\end{equation}
In section \ref{sec.numerical} we solve numerically this system of non linear coupled differential equations for differents value of initial conditions.\\
Nevertheless here we want to study analitically the behavior of the system. In order to do that, we note that $v_{\gamma}$ varies with a characteristic time $\tau_{\gamma} = \frac{C'}{\lambda}$ while $v_{\eta}$ varies with a characteristic time $\tau_{\eta} = \frac{C}{\lambda} $.
Hence if $C' << C$ then $\tau_{\nu} >> \tau_{\gamma}$\footnote{see appendix for numerical estimates.}; in this case $v_{\eta}$ varies very slowly compared to $v_{\gamma}$ and then we can study the evolution of $v_{\gamma}$ in \eqref{eq.velocity_field} considering $v_{\eta}$ as a constant.
This is equivalent to decouple the four equations in \eqref{eq.velocity_field} into two velocity field $\mathbf{f_{\gamma}}$ and $\mathbf{f_{\eta}}$.\\
Let's consider then $\mathbf{f_{\gamma}}$, with $v_{\eta}$ constant:
\begin{equation}
    \label{eq.velocity_field_gamma}
        \begin{cases}
            C' \dot{v_{\gamma}} = - \frac{3}{4} (B'-A') \omega^2 a_k(e)e^k \sin{2\gamma} - \lambda v_{\gamma} + \lambda v_{\eta}\\
            \dot{\gamma} = v_{\gamma}
        \end{cases}
\end{equation}
The equilibrium point is $(\Bar{\gamma}; 0)$, with $\Bar{\gamma}= \frac{1}{2} \arcsin{\left( \frac{4\lambda v_{\eta}}{3 (B'-A') \omega^2 a_k(e)e^k} \right)}$.\\
This equilibrium point exists if and only if 
\begin{equation}
    \label{eq.condition_existence}
     \frac{4\lambda |v_{\eta}|}{3 (B'-A') \omega^2 a_k(e)e^k}  < 1 \implies |v_{\eta}| < \frac{3 (B'-A') \omega^2 a_k(e)e^k}{4\lambda},
\end{equation}
namely if the core is rotating with a velocity not too far from the resonance one.\\
The physical meaning is straightforward. If the friction term $\lambda v_{\eta}$ is bigger than the torque, then the crust stays glued to the core and the two rotate together. If the condition \eqref{eq.condition_existence} is not satisfied, then the solution of the equations, after a small transient, is characterized by a $\gamma$ that grows linearly in time and a $v_{\gamma}$ that oscillates very slowly around the value $v_{\eta}$: $\langle\dot\varphi\rangle = \langle\dot\nu\rangle$.\\
On the other side, if the condition \eqref{eq.condition_existence} is satisfied, then the friction term is weaker than the torque and in this case gravitational attraction of the secondary brings the crust into the spin-orbit resonance (that is the equilibrium point).\\
Notice that in our model the crust will certainly (with probability $1$) reach the equilibrium point if the condition \eqref{eq.condition_existence} is satisfied, which will certainly be verified after a long enough time since $v_{\eta}$ is slowly decreasing due to the presence of friction terms.

Once the crust reaches the resonance, slowly it brings into the resonance also the core. Indeed, if the crust is in resonance, then $v_{\gamma} = 0$, so $\mathbf{f_{\eta}}$ in \eqref{eq.velocity_field} become:
\begin{equation}
    \label{eq.velocity_field_ni}
        \begin{cases}
             C \dot{v_{\eta}} = - \frac{3}{4} (B-A) \omega^2 a_k(e)e^k \sin{2\eta} - (\lambda + \lambda')v_{\eta} -\lambda' \frac{k\omega}{2}\\
            \dot{\eta} = v_{\eta}
        \end{cases}
\end{equation}
In order to simplify the notation, we can rescale the first equation, obtaining:
\begin{equation}
    \label{eq.velocity_field_ni2}
        \begin{cases}
             \dot{v_{\eta}} = - \frac{1}{2} \varepsilon \omega^2 a_k(e)e^k \sin{2\eta} - \frac{(\lambda + \lambda')}{C}v_{\eta} -\frac{\lambda'}{C} \frac{k\omega}{2}\\
            \dot{\eta} = v_{\eta},
        \end{cases}
\end{equation}
with $\varepsilon= \frac{3}{2}\frac{(B-A)}{C}$.\\
The equilibrium point is $(\Bar{\eta}; 0)$, with $\Bar{\eta}= \frac{1}{2} \arcsin{\left( \frac{k \lambda'/C}{3 \varepsilon \omega a_k(e)e^k} \right)}$.\\
This equilibrium point exists if and only if 
\begin{equation}
    \label{eq.condition_existence_ni}
     \frac{2k \lambda'/C}{3 \varepsilon \omega a_k(e)e^k}  < 1 \implies e^k > \frac{2k \lambda'}{3 \varepsilon C \omega a_k(e)},
\end{equation}
namely if the eccentricity is big enough.\\
The equilibrium point $(\Bar{\eta}; 0)$ is asymptotically stable if it exist a Lyapunov function $W(\eta, v_{\eta})$ such that: $W$ has a minimum in $(\Bar{\eta}, 0)$ and $\mathbf{\nabla W} \cdot \mathbf{f_{\eta}} \leq 0$.\\
A Lyapunov function is:
\begin{equation}
    \label{eq.Lyapunov_function}
    W(\eta, v_{\eta}) = \frac{1}{2}v_{\eta}^2 - \frac{1}{4} \varepsilon \omega^2 a_k(e)e^k \cos{2 \eta} + \frac{\lambda'}{C} \frac{k\omega}{2}\eta.
\end{equation}
Indeed:
\begin{equation*}
    \mathbf{\nabla W} \cdot \mathbf{f_{\eta}} = \frac{\partial W}{\partial v_{\eta}} \cdot \Dot{v_{\eta}} + \frac{\partial W}{\partial \eta} \cdot \Dot{\eta} =
\end{equation*}
\begin{equation*}
    = v_{\eta} \left[ -\frac{1}{2} \varepsilon \omega^2 a_k(e) e^k \sin 2\eta - \frac{(\lambda + \lambda')}{C} v_{\eta} - \frac{\lambda'}{C} \frac{k\omega}{2} \right]
\end{equation*}
\begin{equation}
    \label{eq.Lyapunov_proof}
    + \left[ \frac{1}{2} \varepsilon \omega^2 a_k(e)e^k \sin{2 \eta} + \frac{\lambda'}{C} \frac{k\omega}{2} \right] v_{\eta} = -\frac{(\lambda+\lambda')}{C} v_{\eta}^2 \leq 0.
\end{equation}\\
So we can consider the Lyapunov function as an effective energy:
\begin{equation}
    \label{eq.effective_energy}
    E_{eff} = \frac{1}{2} \dot\eta^2 + V_{eff}
\end{equation}
such that
\begin{equation}
    \label{eq.dissipation_energy}
    \frac{dE_{eff}}{dt} = - \frac{(\lambda+\lambda')}{C} \dot\eta^2
\end{equation}
and
\begin{equation}
    \label{eq.effective_potential}
    V_{eff} = - \frac{1}{2} \varepsilon \omega^2 a_k(e)e^k \cos{2 \eta} + \frac{\lambda'}{C} \frac{k\omega}{2}\eta.
\end{equation}\\
Now, following the already mentioned pioneering papers \cite{goldreich1966spin, goldreich1967spin}, we can find a condition which ensure that the equilibrium point is reached. Indeed the core will certainly (with probability $1$) reach the resonance if the energy dissipation $|\Delta E_{eff}|$ between two maxima of the potential (e.g. between $-\frac{\pi}{2}$ and $\frac{\pi}{2}$) is bigger than the corresponding potential variation $|\Delta V_{eff}|$.\\
\begin{figure}
    \centering
    \includegraphics[width=0.6\textwidth]{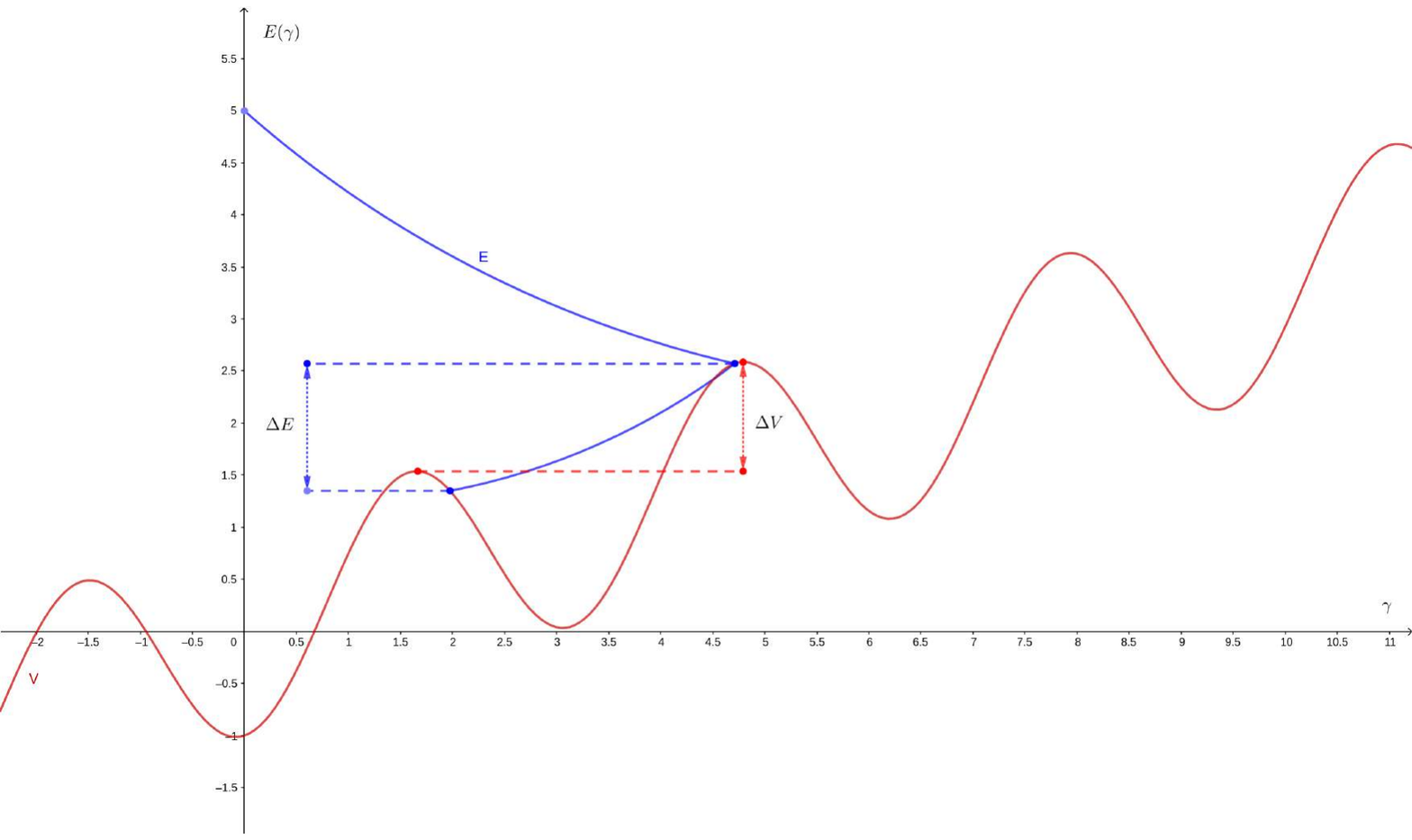}
    \caption{Graphical representation of the condition that ensure the capture in resonance: the energy dissipation |$\Delta E_{eff}$| between two maxima of the potential is bigger than the corresponding potential variation |$\Delta V_{eff}$|. The scale on the $y$-axis has been voluntarily increased.}
    \label{fig.condition}
\end{figure}

|$\Delta V_{eff}$| is easy to determine:
\begin{equation}
    \label{eq.Delta_Veff}
    |\Delta V_{eff}| = \left| V_{eff}\left(\frac{\pi}{2}\right) - V_{eff}\left(-\frac{\pi}{2}\right) \right| = \frac{\lambda'}{C} \frac{k\omega}{2}\pi.
\end{equation}\\
While |$\Delta E_{eff}$| can be determine integrating $\frac{dE_{eff}}{d\eta}$ from $-\frac{\pi}{2}$ to $\frac{\pi}{2}$: 
\begin{equation}
    \label{eq.delta_E_first}
    |\Delta E_{eff}| = \left| \int_{-\frac{\pi}{2}}^{\frac{\pi}{2}} \left( \frac{dE_{eff}}{d\eta} \right) d\eta \right| = \left| \int_{-\frac{\pi}{2}}^{\frac{\pi}{2}} \frac{(\lambda + \lambda')}{C} \dot\eta d\eta \right| = \frac{(\lambda + \lambda')}{C} \int_{-\frac{\pi}{2}}^{\frac{\pi}{2}} \sqrt{2(E_{eff}-V_{eff})} d\eta
\end{equation}
To solve the integral, we can replace $V_{eff}$ with $V>V_{eff}$, obtaining a lower bound for $|\Delta E_{eff}|$. Moreover, as we can see in figure \ref{fig.condition}, we can imagine that the motion reverses its direction near to the maximum of the potential (in this way we still get a lower bound):
\begin{equation*}
    |\Delta E_{eff}| \geq \frac{(\lambda + \lambda')}{C} \int_{-\frac{\pi}{2}}^{\frac{\pi}{2}} \sqrt{2(E_{eff}-V)} d\eta \geq \frac{(\lambda + \lambda')}{C} \int_{-\frac{\pi}{2}}^{\frac{\pi}{2}} \sqrt{2\frac{1}{2} \varepsilon \omega^2 a_k(e)e^k (1+\cos{2\eta})} d\eta = 
\end{equation*}
\begin{equation}
    \label{eq.delta_E_second}
    = \frac{(\lambda + \lambda')}{C} \omega \sqrt{\varepsilon a_k(e)e^k} \int_{-\frac{\pi}{2}}^{\frac{\pi}{2}} \sqrt{1+\cos{2\eta}} d\eta = \frac{(\lambda + \lambda')}{C} \omega \sqrt{8\varepsilon a_k(e)e^k}
\end{equation}
where we have used the relation $1+\cos{2\eta} = 2\cos^2{\eta}$.\\
So, the core will reach the $2(k/2+1) : 2$ spin-orbit resonance with probability $1$ if:
\begin{equation}
    \label{eq.cond_ecc}
    |\Delta E_{eff}| > |\Delta V_{eff}| \implies  \frac{(\lambda + \lambda')}{C} \omega \sqrt{8\varepsilon a_k(e)e^k} > \frac{\lambda'}{C} \frac{k\omega}{2}\pi \implies \frac{\lambda}{\lambda'} > \frac{k \pi }{\sqrt{32\varepsilon a_k(e)e^k}}-1,
\end{equation}
namely if $\lambda$ is bigger enough of $\lambda'$.\\
An analogous computation for the capture in resonance of the crust has been omitted because the linear term in its Lyapunov function has the opposite sign, implying the capture with probability $1$.\\
Notice that in the pioneering works \cite{goldreich1966spin, goldreich1967spin} no assumptions are made about the internal structure of the primary, so in that model the additional viscous dissipation is absent, while it is a fundamental ingredient in our model. We can therefore think that our model generalizes that of Goldreich and Peale, indeed setting $\lambda = 0$ (i.e. neglecting the viscous friction between the two layers) our equation of motion for $\eta$ in \eqref{eq_2equation_coupled} becomes the equation (10) with torque (12a) in \cite{goldreich1966spin}; therefore in this case the problem of a small capture probability arises. On the contrary, if one consider, like we did, also the internal viscous dissipation between the layers of the primary, that problem is overcome: the core is certainly (with probability $1$) captured into the resonance if conditions \eqref{eq.condition_existence_ni} and \eqref{eq.cond_ecc} are satisfied, namely, respectively, it the eccentricity is big enough and if the viscous dissipation is bigger enough with respect the viscoelastic one \footnote{see appendix for numerical estimates.}.

\section{Exit from the resonance}

When the core and the crust are both in resonance, they rotate together with the same angular speed:\\ $\dot\varphi = \dot\nu = \frac{k\omega}{2} \implies v_{\gamma }=v_{\eta} = 0$, so the viscous dissipation in both the equations for $\gamma$ and $\eta$ vanishes.\\
In such a situation the only dissipation source is the viscoelastic term $\lambda' \frac{k\omega}{2}$: 
\begin{equation}
    \label{eq.velocity_field_end}
        \begin{cases}
            C' \dot{v_{\gamma}} = - \frac{3}{4} (B'-A') \omega^2 a_k(e)e^k \sin{2\gamma} \\
            \dot{\gamma} = v_{\gamma} \\
            C \dot{v_{\eta}} = - \frac{3}{4} (B-A) \omega^2 a_k(e)e^k \sin{2\eta} - \lambda' \frac{k\omega}{2}\\
            \dot{\eta} = v_{\eta}
        \end{cases}
\end{equation}
Although the dissipation $\lambda' \frac{k\omega}{2}$ is very small, it slowly tends to circularize the orbit.\\
When the eccentricity becomes so small that condition \eqref{eq.condition_existence_ni} is no longer satisfied, then the system exit from the $2(k/2+1) : 2$ and, while the crust and the core still rotate at the same angular speed, it further decreases until the system reaches, with the same mechanisms, the $2(k'/2+1) : 2$ spin-orbit resonance, with $k'=k-1$.\\
The last but one resonance that the system visits is the $3:2$ spin-orbit resonance($k=1$), as in Sun-Mercury system. The last is, of course, the $1:1$ spin orbit resonance ($k=0$), as in Jupiter-Ganymede system.

\section{Numerical simulations}\label{sec.numerical}
\begin{figure}
    \centering
    \includegraphics[width=0.4\textwidth]
                    {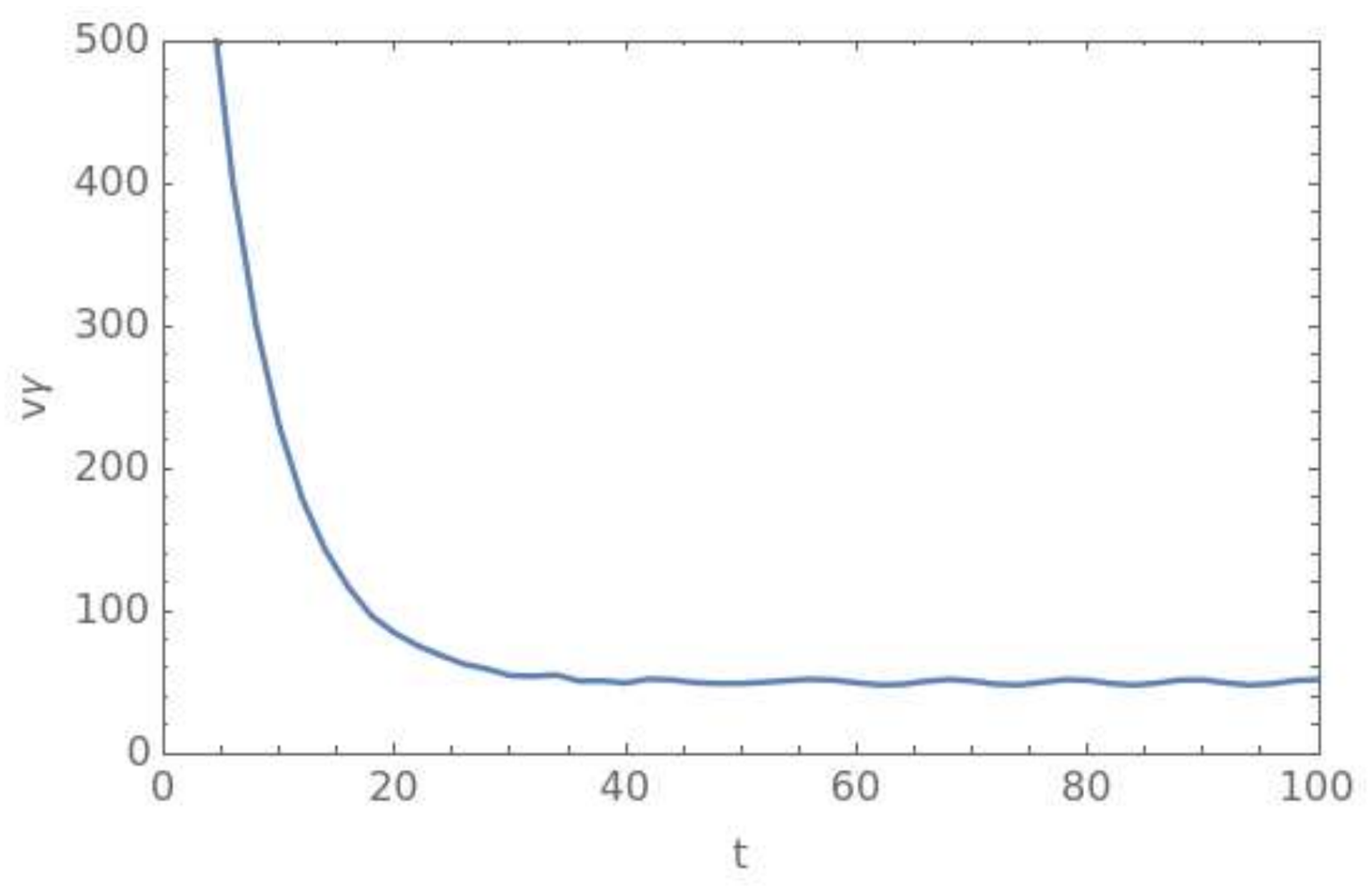}
                    \hspace{0.1\textwidth}
    \includegraphics[width=0.4\textwidth]
                    {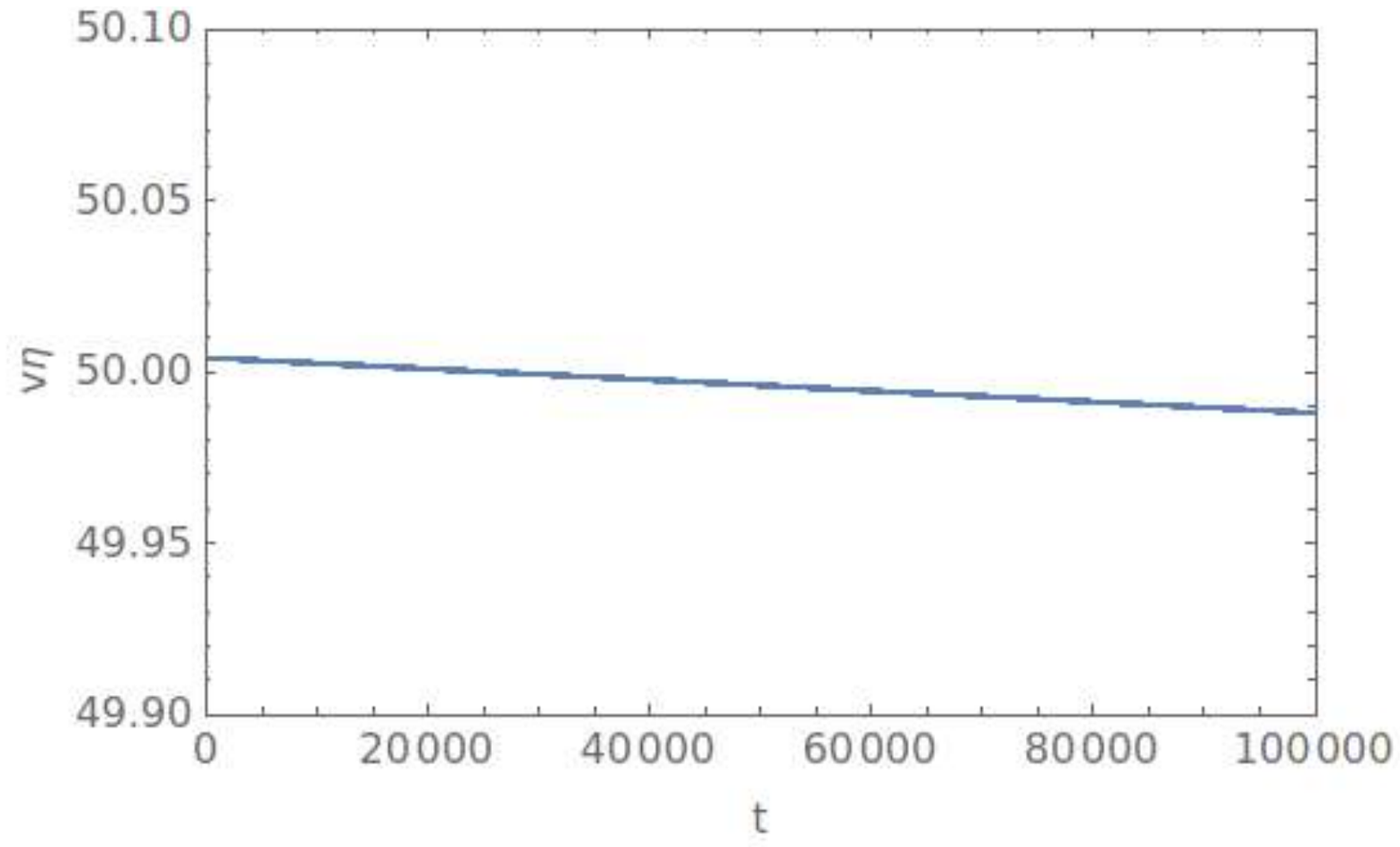}
    \caption
    {
    Numerical solution for the system with initial conditions $(\gamma; v_{\gamma}; \eta; v_{\eta}) = (0.1; 1000\,\frac{1}{y}; 0.1; 50\,\frac{1}{y})$, i.e. when  $v_{\eta}$ does not satisfy condition \eqref{eq.condition_existence}.\\
    In the picture on the left we have plotted $v_{\gamma}$ vs time for $100$ time steps. As we can see, after about $30$ time steps $v_{\gamma}$ reaches the value of $v_{\eta}$ and oscillates around it.\\
    In the picture on the right we have plotted $v_{\eta}$ vs time for $100000$ time steps. As we can see, $v_{\eta}$ remain almost constant: indeed he decreases very slowly on a very long time scale.
    }
    \label{fig.non_capture}
\end{figure}

\begin{figure}
    \centering
    \includegraphics[width=0.4\textwidth]
                    {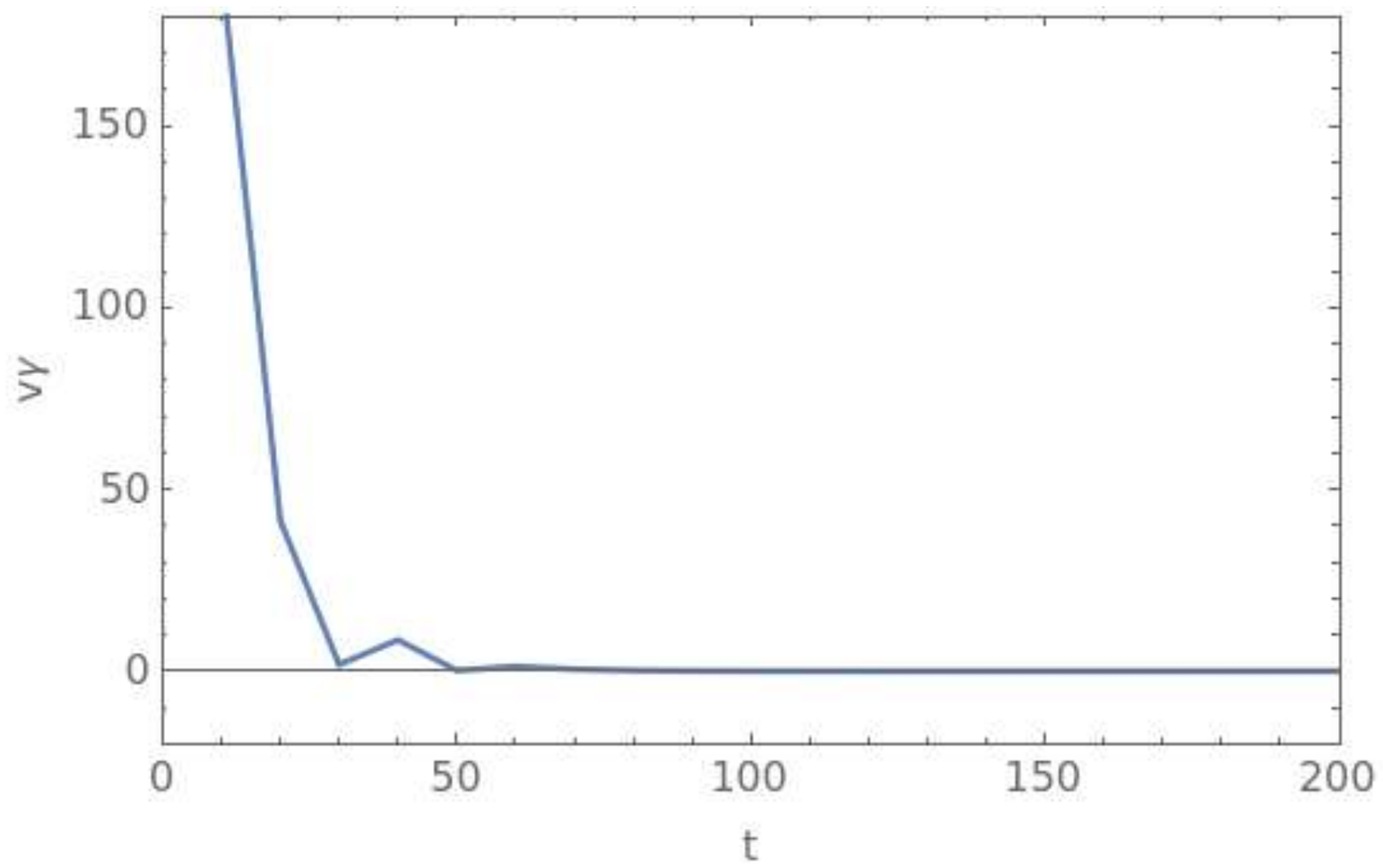}
                    \hspace{0.1\textwidth}
    \includegraphics[width=0.4\textwidth]
                    {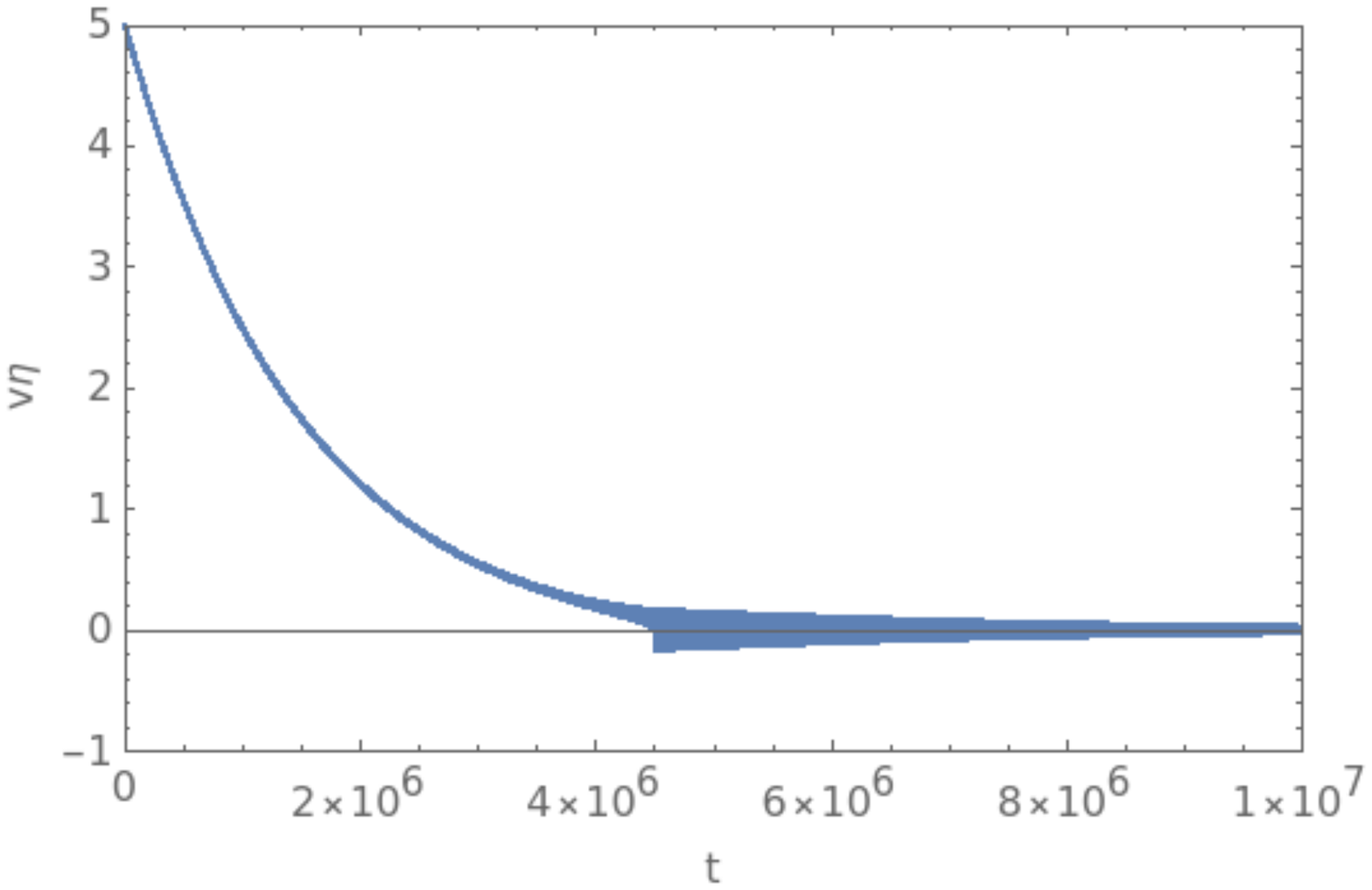}
    \caption
    {
    Numerical solution for the system with initial conditions $(\gamma; v_{\gamma}; \eta; v_{\eta}) = (0.1; 1000\,\frac{1}{y}; 0.1; 5\,\frac{1}{y})$, i.e. when  $v_{\eta}$ satisfies condition \eqref{eq.condition_existence}.\\
    In the picture on the left we have plotted $v_{\gamma}$ vs time for $200$ time steps. As we can see, after about $50$ time steps $v_{\gamma}$ vanishes and oscillates around zero.\\
    In the picture on the right we have plotted $v_{\eta}$ vs time for $10^7$ time steps. As we can see, $v_{\eta}$ vanishes but on a much longer time scale (on the order of $10^6$ iterations).
    }
    \label{fig.capture}
\end{figure}

In this section we present some preliminary numerical simulations integrating the set of coupled equations \eqref{eq.velocity_field} using Mathematica.\\
The purpose of these simulations is to qualitatively observe the dynamics of the system and to note the presence of different time scales in order to justify the assumptions made before regarding the way to decouple the equations.\\
More in-depth simulations will be the subject of future works, as well as simulations regarding the orbital dynamics of systems in $1:1$ spin-orbit resonance. The aim is to observe how the internal structure modeled in this paper can influence the evolution of systems in $1:1$ spin orbit resonance, such as the Jupiter-Ganymede system. This could provide a support framework for the JUICE mission to analyze the data collected by the space craft.\\
In order to perform the numerical integration in a more handy way, we rewrite equations \eqref{eq.velocity_field} as:
\begin{equation}
    \label{eq.velocity_field_4}
        \begin{cases}
            \dot{v_{\gamma}} = - \frac{3}{4} \varepsilon' \omega^2 a_k(e)e^k \sin{2\gamma} - \frac{1}{\tau_{\gamma}} (v_{\gamma} -v_{\eta})\\
            \dot{\gamma} = v_{\gamma}\\
            \dot{v_{\eta}} = - \frac{3}{4} \varepsilon \omega^2 a_k(e)e^k \sin{2\eta} + \frac{1}{\tau_{\eta}} (v_{\gamma} - v_{\eta}) - \frac{1}{\tau'_{\eta}} (v_{\eta} + \frac{k\omega}{2})\\
            \dot{\eta} = v_{\eta}
        \end{cases}
\end{equation}
with: $\tau_{\gamma} = \frac{C'}{\lambda}$, $\tau_{\eta} = \frac{C}{\lambda}$ and $\tau'_{\eta} = \frac{C}{\lambda'}$.\\
For the purpose of this work, we focus our simulations on a particular spin orbit resonance, namely the 3:2. This correspond to set $k=1$ and consequently $a_k(e)$ evaluated at $e=0$ is $a_1(0) = 3$; the exploration of the cascade in subsequent spin orbit resonances will be the subject of future works.\\
Finally, in order to obtain plausible simulations, we consider the parameters values of Ganymede, summarized in table \ref{tab.parameters}.\\
The equations become the following:
\begin{equation}
    \label{eq.velocity_field_5}
        \begin{cases}
            \dot{v_{\gamma}} = - 202,5 \sin{2\gamma} - 0,167 (v_{\gamma} -v_{\eta})\\
            \dot{\gamma} = v_{\gamma}\\
            \dot{v_{\eta}} = - 0,0135 \sin{2\eta} + 6,67 \times 10^{-7} (v_{\gamma} - v_{\eta}) - 6,67 \times 10^{-10} (v_{\eta} + 150)\\
            \dot{\eta} = v_{\eta}
        \end{cases}
\end{equation}
where we have used S.I. units except for time, which is expressed in years.\\ 
We first integrated the equations for $10^5$ iterations with initial conditions\\ $(\gamma; v_{\gamma}; \eta; v_{\eta})=(0.1; 1000\,\frac{1}{y}; 0.1; 50\,\frac{1}{y})$, i.e. when  $v_{\eta}$ does not satisfy condition \eqref{eq.condition_existence}; the result are shown in figure \ref{fig.non_capture}. As we can see, in this case $v_{\gamma}$ decrease on a very short time until he reaches the value of $v_{\eta}$ and from then it oscillates around this value (left picture in figure \ref{fig.non_capture}) never approaching zero. $v_{\eta}$, on the contrary, remain almost constant: indeed it decreases very slowly on a very long time scale (right picture in figure \ref{fig.non_capture}).\\
Then we integrated the equations for $10^7$ iterations with initial conditions\\ $(\gamma; v_{\gamma}; \eta; v_{\eta}) = (0.1; 1000\,\frac{1}{y}; 0.1; 5\,\frac{1}{y})$, i.e. when $v_{\eta}$ satisfies condition \eqref{eq.condition_existence}; the result are shown in figure \ref{fig.capture}. As we can see, in this case $v_{\gamma}$ vanishes on a very short time and from then it oscillates around zero (left picture in figure \ref{fig.capture}). $v_{\eta}$, on the contrary, also vanishes but on a longer time scale (right picture in figure \ref{fig.capture}).\\
At some point the system will exit from resonance. This process occurs in a very long time scales\footnote{see again appendix for numerical estimates.}. In this paper we are not interested in the numerical investigation of this process, that will be the subject of future works.

\appendix

\section{Astronomical estimates}
In this section we want to propose a possible way to estimate some quantities involved in our model from values know in literature, summarized in table \ref{tab.parameters}.\\
To this end let's consider the equation for $v_{\eta}$ in \eqref{eq.velocity_field}:
\begin{equation*}
    C \dot{v_{\eta}} = - \frac{3}{4} (B-A) \omega^2 a_k(e)e^k \sin{2\eta} + \lambda (v_{\gamma} - v_{\eta}) -\lambda' (v_{\eta} + \frac{k\omega}{2})
\end{equation*}
\begin{equation}
    \label{eq.veta_appendix}
    = - \frac{3}{4} (B-A) \omega^2 a_k(e)e^k \sin{2\eta} + T_v +T_t,
\end{equation} 
with:
$T_v = \lambda (v_{\gamma} - v_{\eta})$ the viscous torque and $T_t = -\lambda' (v_{\eta} +\frac{k\omega}{2})$ the viscoelastic tidal torque.\\
We can compute $T_v$ considering a two-layer body having a solid core that rotates with angular speed $v_{\nu}$ and a solid crust that rotates with angular speed $v_{\varphi}$, separated by a fluid (i.e. an ocean) of depth $h$. Hence the crust rotates with velocity $v_{\nu}-v_{\varphi} = v_{\eta}-v_{\gamma}$ with respect to the core.\\
Finally let's suppose that the viscous friction force is $\eta$ times the gradient of velocity (here $\eta$ represents the viscosity of the liquid). Hence the friction torque with respect the rotational axis is:
\begin{equation}
    \label{eq.torque_1}
    |T_v|= \int_0^{2\pi} \int_0^{\pi} \frac{\eta (v_{\eta}-v_{\gamma}) r}{h} r R^2 \sin{\vartheta} d\vartheta d\phi,
\end{equation}
with $r=R \sin{\vartheta}$ the distance between the rotational axis ($z$-axis) and the point of geographic coordinate $(\phi, \vartheta)$.\\
\begin{equation}
    \label{eq.torque_2}
    |T_v|= 2\pi \frac{\eta (v_{\eta}-v_{\gamma}) R^4}{h} \int_0^{\pi} \sin^3{\vartheta} d\vartheta = \frac{8\pi}{3} \frac{\eta (v_{\eta}-v_{\gamma}) R^4}{h}.
\end{equation}
So, the friction coefficient $\lambda$ used from \eqref{eq_2equation_coupled} onwards is given by:
\begin{equation}
    \label{eq.lambda_expression}
    \lambda = \frac{8\pi}{3} \frac{\eta R^4}{h}.
\end{equation}


On the other hand, we can compute $T_t$ from the MacDonald's tidal torque formula:
\begin{equation}
    \label{eq.tidal_torque_1}
    |T_t|= \frac{3}{2}\frac{k_2}{Q}\frac{Gm^2R^5}{a^6}.
\end{equation}
Hence, if we suppose $v_{\eta}< \omega$:
\begin{equation}
    \label{eq.lambdap_expression}
    \lambda'= \frac{3}{k\omega}\frac{k_2}{Q}\frac{Gm^2R^5}{a^6}.
\end{equation}
Finally:
\begin{equation}
    \frac{\lambda}{\lambda'} = \frac{8\pi \eta a^6 Q k \omega}{9 h k_2 G m^2 R}
\end{equation}\\
Considering the values of the parameters known in literature (table \ref{tab.parameters}), we obtain the estimate of $\lambda/\lambda'$ for Ganymede and Mercury are:
\begin{equation}
    \left(\frac{\lambda}{\lambda'}\right)_G \simeq 10^3; \text{ \ \ \ \ \ } \left(\frac{\lambda}{\lambda'}\right)_M \simeq 10^{16}.
\end{equation}\\
If we now consider condition \eqref{eq.cond_ecc} for $k=1$ ($3:2$ spin-orbit resonance), we can see that for Mercury's parameters it is certainly satisfied. Whereas in the case of Ganymede, the condition require an $\lambda/\lambda'$ ratio exactly of the same order as the one we estimate using current values for the parameters, so is not obvious that Ganymede could have crossed the $3:2$ spin-orbit resonance. Since some of Ganymede's parameters are not known exactly, but in a range, the values of $\lambda/\lambda'$ ratio could be update in future with more accurate measurements.\\
Finally, the estimate value $\left(\frac{\lambda}{\lambda'}\right)_M \simeq 10^{16}$ is a purely indicative lower bound: indeed in the case of Mercury we expect that $\lambda'$ is bigger than the value we found since the molten mantle too gives an important contribution to the dissipation. Nevertheless an in-deep study of this aspect is beyond the aim of this work.

\section{Characteristic time scales}
In this section we want to estimate the characteristic time scales of the system in order to justify all the assumptions we made regarding the decoupling of equations \eqref{eq.velocity_field}:
\begin{equation}
    \label{eq.velocity_field_3}
        \begin{cases}
            C' \dot{v_{\gamma}} = - \frac{3}{4} (B'-A') \omega^2 a_k(e)e^k \sin{2\gamma} - \lambda (v_{\gamma} -v_{\eta})\\
            \dot{\gamma} = v_{\gamma}\\
            C \dot{v_{\eta}} = - \frac{3}{4} (B-A) \omega^2 a_k(e)e^k \sin{2\eta} + \lambda (v_{\gamma} - v_{\eta}) -\lambda' (v_{\eta} + \frac{k\omega}{2})\\
            \dot{\eta} = v_{\eta}
        \end{cases}
\end{equation}
As we have seen in the paper, we can identify three well-separated processes: the entrance in resonance for the crust, the entrance in resonance for the core and the exit from resonance.\\
We will see that these three processes occur on very different time scales.
Indeed, the entrance in resonance for the crust occurs in a characteristic time $\tau_{\gamma} = \frac{C'}{\lambda}$ (characteristic time of the dynamics of $v_{\gamma}$).\\
On the other hand, the entrance in resonance for the core occurs in a characteristic time $\tau_{\eta} = \frac{C}{\lambda}$ (characteristic time of the dynamics of $v_{\eta}$).\\
Finally, the exit from resonance occurs in a characteristic time $\tau_{\eta}' = \frac{C}{\lambda'}$ (characteristic time of the dynamics of $v_{\eta}$ in equation \eqref{eq.velocity_field_end}, that coincides with the characteristic time of the dynamics of the eccentricity).\\
Since $C' << C$ and $\lambda' << \lambda$, then: $\tau_{\gamma} << \tau_{\eta} << \tau_{\eta}'$ and thus the three processes mentioned above have a well-separated time scales.\\
Since $C \propto R^5$, it is reasonable to assume that $C'/C$ of the order $10^{-3}$ or $10^{-4}$: indeed the thickness of the crust is about $10^4\,m$ or $10^5\,m$ while the radius of the core is about $10^6\,m$ for both Mercury and Ganymede.\\  
The estimate of $\tau_{\gamma}; \tau_{\eta}; \tau_{\eta}'$ for Ganymede are:
\begin{equation}
    \tau_{\gamma} \simeq 10^9\,s; \text{ \ \ \ \ \ } \tau_{\eta} \simeq 10^{12}\,s; \text{ \ \ \ \ \ } \tau_{\eta}' \simeq 10^{15}\,s.
\end{equation} 
We note that $\tau_{\eta}'$, that is the characteristic time of exit from resonance, coincides with the characteristic time of variation of eccentricity. Our value is in good agreement with that known in literature:\\
$10^8\,y \simeq 10^{15}\,s$, see, for instance \cite{MurrayDermott, showman1997coupled}.

\begin{table}
    \centering
        \begin{tabular}{|c c c|} 
    \hline
    & Table of physical parameters values & \\
    \hline
     & Ganymede & Mercury \\ [1ex] 
    \hline
    Mean motion $(\omega)$ & $1.0 \times 10^{-5}$ & $8.3 \times 10^{-7}$ \\ 
    
    Semi-major axis $(a)$ & $1.1 \times 10^9$ & $5.8 \times 10^{10}$ \\
    
    Eccentricity $(e)$ & $1.3 \times 10^{-3}$ & $2.1 \times 10^{-1}$ \\
    
    Mean radius $(R)$ & $2.6 \times 10^6$ & $2.4 \times 10^6$ \\ 
    
    Mass $(m)$ & $1.5 \times 10^{23}$ & $3.3 \times 10^{23}$ \\
    
    Equatorial ellipticity $(\varepsilon)$ & $(10^{-4})$ & $(10^{-4})$ \\
    
    Tidal quality factor $(Q)$ & $(100)$ & $(100)$ \\
    
    Tidal Love number $(K_2)$ & $(0.02)$ & $(0,1)$ \\ 
    
    Viscosity of "mantle" $(\eta)$ & $1.6 \times 10^{-3}$ & $(10^3)$ \\
    
    Depth of "mantle" $(h)$ & $ (10^5) $ & $(4 \times 10^5)$ \\[1ex] 
    \hline
\end{tabular}    
    \caption{Physical parameters values expressed with two significant digits in SI units. The values in round brackets are the most recent estimate known in literature.}
    \label{tab.parameters}
\end{table}

\section*{Acknowledgements}
We benefited of several comments by Christoph Lothka and Giuseppe Pucacco.\\
BS acknowledges the support of the Italian MIUR Department of Excellence grant (CUP E83C18000100006).\\
MV has been supported through the ASI Contract n.2023-6-HH.0, Scientific Activities for JUICE, E phase (CUP F83C23000070005).

\printbibliography

@book{MurrayDermott,
  title={Solar system dynamics},
  author={Murray, Carl D and Dermott, Stanley F},
  year={1999},
  publisher={Cambridge university press}
}

@article{efroimsky2013tidal,
  title={Tidal friction and tidal lagging. Applicability limitations of a popular formula for the tidal torque},
  author={Efroimsky, Michael and Makarov, Valeri V},
  journal={The Astrophysical Journal},
  volume={764},
  number={1},
  pages={26},
  year={2013},
  publisher={IOP Publishing}
}

@article{efroimsky2015tidal,
  title={Tidal evolution of asteroidal binaries. Ruled by viscosity. Ignorant of rigidity},
  author={Efroimsky, Michael},
  journal={The Astronomical Journal},
  volume={150},
  number={4},
  pages={98},
  year={2015},
  publisher={IOP Publishing}
}

@article{ferraz2015dissipative,
  title={Dissipative forces in celestial mechanics. 30o Col{\'o}quio Brasileiro de Matem{\'a}tica},
  author={Ferraz-Mello, S and Grotta-Ragazzo, C and dos Santos, L Ruiz},
  journal={Publica{\c{c}}oes Matem{\'a}ticas, IMPA},
  year={2015}
}

@article{scoppola2022tides,
  title={Tides and dumbbell dynamics},
  author={Scoppola, Benedetto and Troiani, Alessio and Veglianti, Matteo},
  journal={Regular and Chaotic Dynamics},
  volume={27},
  number={3},
  pages={369--380},
  year={2022},
  publisher={Springer}
}

@article{scoppola2022shaken,
  title={Shaken dynamics on the 3d cubic lattice},
  author={Scoppola, Benedetto and Troiani, Alessio and Veglianti, Matteo},
  journal={Electronic Journal of Probability},
  volume={27},
  pages={1--26},
  year={2022},
  publisher={The Institute of Mathematical Statistics and the Bernoulli Society}
}

@article{apollonio2022shaken,
  title={Shaken dynamics: an easy way to parallel Markov Chain Monte Carlo},
  author={Apollonio, Valentina and D’Autilia, Roberto and Scoppola, Benedetto and Scoppola, Elisabetta and Troiani, Alessio},
  journal={Journal of Statistical Physics},
  volume={189},
  number={3},
  pages={39},
  year={2022},
  publisher={Springer}
}

@article{apollonio2022metastability,
  title={Metastability for the Ising model on the hexagonal lattice},
  author={Apollonio, Valentina and Jacquier, Vanessa and Nardi, Francesca Romana and Troiani, Alessio},
  journal={Electronic Journal of Probability},
  volume={27},
  pages={1--48},
  year={2022},
  publisher={The Institute of Mathematical Statistics and the Bernoulli Society}
}

@article{d2021parallel,
  title={Parallel simulation of two-dimensional Ising models using probabilistic cellular automata},
  author={D’Autilia, Roberto and Andrianaivo, Louis Nantenaina and Troiani, Alessio},
  journal={Journal of Statistical Physics},
  volume={184},
  pages={1--22},
  year={2021},
  publisher={Springer}
}

@article{apollonio2019criticality,
  title={Criticality of measures on 2-d Ising configurations: from square to hexagonal graphs},
  author={Apollonio, Valentina and D’Autilia, Roberto and Scoppola, Benedetto and Scoppola, Elisabetta and Troiani, Alessio},
  journal={Journal of Statistical Physics},
  volume={177},
  number={5},
  pages={1009--1021},
  year={2019},
  publisher={Springer}
}

@article{goldreich1966spin,
  title={Spin-orbit coupling in the solar system},
  author={Goldreich, Peter and Peale, Stanton},
  journal={Astronomical Journal, Vol. 71, p. 425 (1966)},
  volume={71},
  pages={425},
  year={1966}
}

@article{goldreich1967spin,
  title={Spin-orbit coupling in the solar system. II. The resonant rotation of Venus},
  author={Goldreich, Peter and Peale, Stanton},
  journal={Astronomical Journal, Vol. 72, p. 662 (1967)},
  volume={72},
  pages={662},
  year={1967}
}

@article{noyelles2014spin,
  title={Spin--orbit evolution of Mercury revisited},
  author={Noyelles, Benoit and Frouard, Julien and Makarov, Valeri V and Efroimsky, Michael},
  journal={Icarus},
  volume={241},
  pages={26--44},
  year={2014},
  publisher={Elsevier}
}

@article{corbi,
  title={Seismic variability of subduction thrust faults: Insights from laboratory models},
  author={Corbi, F. and Funiciello, . and Faccenna, C. and Ranalli, G. and  Heuret, A.},
  journal={Journal of Geophysical Research},
  volume={116},
  pages={1--14},
  year={2011},
  publisher={American Geophysical Union}
}

@article{smith2012gravity,
  title={Gravity field and internal structure of Mercury from MESSENGER},
  author={Smith, David E and Zuber, Maria T and Phillips, Roger J and Solomon, Sean C and Hauck, Steven A and Lemoine, Frank G and Mazarico, Erwan and Neumann, Gregory A and Peale, Stanton J and Margot, Jean-Luc and others},
  journal={science},
  volume={336},
  number={6078},
  pages={214--217},
  year={2012},
  publisher={American Association for the Advancement of Science}
}

@article{gomez2022gravity,
  title={Gravity Field of Ganymede after the Juno Extended Mission},
  author={Gomez Casajus, Luis and Ermakov, AI and Zannoni, M and Keane, JT and Stevenson, D and Buccino, DR and Durante, D and Parisi, M and Park, RS and Tortora, P and others},
  journal={Geophysical Research Letters},
  volume={49},
  number={24},
  pages={e2022GL099475},
  year={2022},
  publisher={Wiley Online Library}
}

@article{chen2021exponential,
  title={Exponential stability of fast driven systems, with an application to celestial mechanics},
  author={Chen, Qinbo and Pinzari, Gabriella},
  journal={Nonlinear Analysis},
  volume={208},
  pages={112306},
  year={2021},
  publisher={Elsevier}
}

@article{baland2019coupling,
  title={Coupling between the spin precession and polar motion of a synchronously rotating satellite: application to Titan},
  author={Baland, Rose-Marie and Coyette, Alexis and Van Hoolst, Tim},
  journal={Celestial Mechanics and Dynamical Astronomy},
  volume={131},
  pages={1--50},
  year={2019},
  publisher={Springer}
}

@article{celletti2000hamiltonian,
  title={Hamiltonian stability of spin--orbit resonances in celestial mechanics},
  author={Celletti, Alessandra and Chierchia, Luigi},
  journal={Celestial Mechanics and Dynamical Astronomy},
  volume={76},
  pages={229--240},
  year={2000},
  publisher={Springer}
}

@article{calleja2022kam,
  title={KAM quasi-periodic tori for the dissipative spin--orbit problem},
  author={Calleja, Renato and Celletti, Alessandra and Gimeno, Joan and de la Llave, Rafael},
  journal={Communications in Nonlinear Science and Numerical Simulation},
  volume={106},
  pages={106099},
  year={2022},
  publisher={Elsevier}
}

@article{correia2018effects,
  title={The effects of deformation inertia (kinetic energy) in the orbital and spin evolution of close-in bodies},
  author={Correia, ACM and Ragazzo, C and Ruiz, LS},
  journal={Celestial Mechanics and Dynamical Astronomy},
  volume={130},
  pages={1--30},
  year={2018},
  publisher={Springer}
}

@article{ragazzo2022librations,
  title={Librations of a body composed of a deformable mantle and a fluid core},
  author={Ragazzo, Clodoaldo and Bou{\'e}, Gwena{\"e}l and Gevorgyan, Yeva and Ruiz, Lucas S},
  journal={Celestial Mechanics and Dynamical Astronomy},
  volume={134},
  number={2},
  pages={10},
  year={2022},
  publisher={Springer}
}

@article{lari2022orbit,
  title={Orbit determination methods for interplanetary missions: development and use of the Orbit14 software},
  author={Lari, Giacomo and Schettino, Giulia and Serra, Daniele and Tommei, Giacomo},
  journal={Experimental Astronomy},
  volume={53},
  number={1},
  pages={159--208},
  year={2022},
  publisher={Springer}
}

@article{lari2018semi,
  title={A semi-analytical model of the Galilean satellites’ dynamics},
  author={Lari, Giacomo},
  journal={Celestial Mechanics and Dynamical Astronomy},
  volume={130},
  number={8},
  pages={50},
  year={2018},
  publisher={Springer}
}

@article{correia2009mercury,
  title={Mercury's capture into the 3/2 spin--orbit resonance including the effect of core--mantle friction},
  author={Correia, Alexandre CM and Laskar, Jacques},
  journal={Icarus},
  volume={201},
  number={1},
  pages={1--11},
  year={2009},
  publisher={Elsevier}
}

@article{correia2010long,
  title={Long-term evolution of the spin of Mercury: I. Effect of the obliquity and core--mantle friction},
  author={Correia, Alexandre CM and Laskar, Jacques},
  journal={Icarus},
  volume={205},
  number={2},
  pages={338--355},
  year={2010},
  publisher={Elsevier}
}

@article{antognini2014spin,
  title={The spin--orbit resonances of the Solar System: a mathematical treatment matching physical data},
  author={Antognini, Francesco and Biasco, Luca and Chierchia, Luigi},
  journal={Journal of Nonlinear Science},
  volume={24},
  pages={473--492},
  year={2014},
  publisher={Springer}
}

@article{bartuccelli2015high,
  title={The high-order Euler method and the spin--orbit model. A fast algorithm for solving differential equations with small, smooth nonlinearity},
  author={Bartuccelli, Michele V and Deane, Jonathan HB and Gentile, Guido},
  journal={Celestial Mechanics and Dynamical Astronomy},
  volume={121},
  pages={233--260},
  year={2015},
  publisher={Springer}
}

@article{folonier2017tidal,
  title={Tidal synchronization of an anelastic multi-layered body: Titan’s synchronous rotation},
  author={Folonier, Hugo A and Ferraz-Mello, Sylvio},
  journal={Celestial Mechanics and Dynamical Astronomy},
  volume={129},
  pages={359--396},
  year={2017},
  publisher={Springer}
}

@article{showman1997coupled,
  title={Coupled orbital and thermal evolution of Ganymede},
  author={Showman, Adam P and Stevenson, David J and Malhotra, Renu},
  journal={Icarus},
  volume={129},
  number={2},
  pages={367--383},
  year={1997},
  publisher={Elsevier}
}
\end{document}